\documentclass[11pt]{article}

\usepackage[utf8]{inputenc} %
\usepackage[T1]{fontenc}    %
\usepackage[table]{xcolor}  %
\definecolor{forest}  {rgb}{0,.4,0} 
\definecolor{midnight}  {rgb}{0,0,.5} 
\usepackage{hyperref}
\hypersetup{colorlinks=true,urlcolor=midnight,citecolor=forest}
\usepackage{graphicx}
\usepackage{url}            %
\usepackage{booktabs}       %
\usepackage{amsfonts}       %
\usepackage{nicefrac}       %
\usepackage{microtype}      %
\usepackage{natbib}
\usepackage{multirow}
\usepackage{tikz,pgfplots} %
\pgfplotsset{compat=1.18}
\usetikzlibrary{arrows,calc,math,decorations.pathreplacing,positioning,shapes,shapes.geometric,fit,backgrounds,external}
\usepackage[labelfont=bf]{caption}
\usepackage{subcaption}
\usepackage{rotating}
\usepackage{mathtools}
\usepackage{svg}
\usepackage{listings}
\usepackage[linesnumbered,boxed,algoruled,noline,noend]{algorithm2e}
\usepackage{pdflscape}
\let\oldnl\nl%
\newcommand{\nonl}{\renewcommand{\nl}{\let\nl\oldnl}}%

\usepackage[normalem]{ulem}

\SetCommentSty{mycommfont}

\DeclarePairedDelimiter{\norm}{\lVert}{\rVert}

\DeclareMathOperator*{\argmin}{argmin}
\DeclareMathOperator*{\diag}{diag}

\newcommand{\R}{\mathbb{R}}
\newcommand{\scat}{\mathrm{scat}}
\newcommand{\inn}{\mathrm{in}}

\newcommand{\cF}{\mathcal{F}}
\newcommand{\uin}{u_{\inn}}
\newcommand{\usc}{u_{\scat}}
\newcommand{\utot}{u}
\newcommand{\kbar}{\nicefrac{k}{2\pi}}
\newcommand{\jacob}{\mathrm{D}}
\newcommand{\DF}{\jacob\mathcal{F}}
\newcommand{\gennn}{\mathrm{NN}} %
\newcommand{\gme}{g} %

\usepackage[capitalize,noabbrev]{cleveref}
\usepackage[letterpaper,top=2cm,bottom=2cm,left=3cm,right=3cm,marginparwidth=1.75cm]{geometry}
\usepackage{bm}

\newcommand{\mmgunet}{{HPS-CNN}}
\newcommand{\hpscnn}{{\textrm{HPS-CNN}}}

\usepackage[mode=buildnew]{standalone}
\newcolumntype{L}{>{$}l<{$}}
\newcolumntype{C}{>{$}c<{$}}
\newcolumntype{R}{>{$}r<{$}}
\usepackage{makecell}
\pgfplotsset{
    compat = 1.18,
    barplot/.style = {
        very thick,
        error bars/.cd,
        y dir=both,
        y explicit,
        error bar style={semithick},
    },
    errorplot/.style = {
        very thick,
        every mark/.append style=solid,
        error bars/.cd,
        y dir=both,
    },
}
\usepackage{authblk}

\title{A Model-Guided Neural Network Method for the Inverse Scattering Problem}
\date{}
\author[1]{Olivia Tsang}
\author[2]{Owen Melia}
\author[3,4]{Vasileios Charisopoulos}
\author[3,6]{Jeremy Hoskins}
\author[5,6]{Yuehaw Khoo}
\author[1,3,5,6]{Rebecca Willett}
\affil[1]{Department of Computer Science, University of Chicago}
\affil[2]{Center for Computational Mathematics, Flatiron Institute}
\affil[3]{National Institute for Theory and Mathematics in Biology}
\affil[4]{Department of Electrical \& Computer Engineering, University of Washington}
\affil[5]{Data Science Institute, University of Chicago}
\affil[6]{Computational and Applied Mathematics, Department of Statistics, University of Chicago}

\begin{document}
\maketitle

\begin{abstract}

Inverse medium scattering is an ill-posed, nonlinear wave-based imaging problem arising in medical imaging, remote sensing, and non-destructive testing.
Machine learning (ML) methods offer increased inference speed and flexibility in capturing prior knowledge of imaging targets relative to classical optimization-based approaches; however, they perform poorly in regimes where the scattering behavior is highly nonlinear.
A key limitation is that ML methods struggle to incorporate the physics governing the scattering process, which are typically inferred implicitly from the training data or loosely enforced via architectural design.
In this paper, we present a method that endows a machine learning framework with explicit knowledge of problem physics, in the form of a differentiable solver representing the forward model.
The proposed method progressively refines reconstructions of the scattering potential using measurements at increasing wave frequencies, following a classical strategy to stabilize recovery.
Empirically, we find that our method provides high-quality reconstructions at a fraction of the computational or sampling costs of competing approaches.

\end{abstract}
\section{Introduction}

Wave scattering is an imaging method with applications spanning medicine, non-destructive testing and remote sensing.
The goal in these applications is to image the interior of an inhomogeneous medium, sometimes called the \emph{scattering potential},
which is probed by a set of incident waves at different frequencies that are scattered at spatially varying speeds;
in particular, the structure of the scattered wave field outside the area of interest characterizes the interior of the medium itself.
Reconstructing the scattering potential from observations of the scattered waves is known as the \emph{inverse scattering problem}. This task 
is computationally challenging, as the
physical model (i.e., the \emph{forward model}) governing the wave scattering process is described by
the solution of an elliptic partial differential equation (PDE), resulting in an ill-posed, nonlinear inverse problem~\citep{colton_inverse_2019}.

Classical approaches use optimization methods to find an estimated scattering potential that is both a good fit to the observations and well aligned with prior knowledge of the scattering potential. For instance, we might model the scattering potential as having a small norm or being well-represented in a band-limited basis. More recent methods for solving inverse problems using machine learning leverage training data to learn to solve the inverse problem, which can both (a) reduce the amount of computation needed for each new scattering potential reconstruction and (b) effectively learn a data-based regularizer, which can yield smaller errors and increased robustness to noise in ill-posed settings. These ideas have been widely explored in the context of \textit{linear} inverse problems \citep{arridge_solving_2019,ongie_deep_2020,meinhardt_learning_2017,aggarwal_modl_2019,gilton_neumann_2020,gong_learning_2020}. However, when the forward model exhibits strongly nonlinear behavior, the applicability and efficacy of these methods can be limited \citep{melia_multi-frequency_2025}.

Our goal is to learn an operator that quickly maps observed scattered wave measurements to accurate estimates of the scattering potential. 
Our approach will incorporate both physical knowledge of the forward model  and prior knowledge about the target scattering potentials learned from training data.
To this end, we propose embedding the forward model---in the form of a differentiable PDE solver---in a neural network reconstruction method. At a high level, our method alternates between calculating the gradient of a natural nonlinear least-squares objective, which is evaluated using the PDE solver, and applying learned filtering operations with 2D convolutional networks.
The structure resembles that of the algorithm unrolling \citep{aggarwal_modl_2019} or learning to optimize \citep{li_learning_2016} frameworks, but, crucially, each step uses measurements taken from higher wave frequency than the last; consequently, each step acts on a different objective function, akin to homotopy methods \citep{dunlavy_homotopy_2005,watson_modern_1989}. Our approach is inspired by the  \textit{recursive linearization} algorithm \citep{chen_recursive_1995}, which targets a sequence of nonlinear least-squares problems corresponding to increasing incident wave frequencies, resulting in a sequence of intermediate reconstructions; each of these reconstructions serves as the initial guess for the next least-squares problem.
We refer to this pattern as \textit{progressive refinement}, since each reconstruction contains higher spatial frequency content than the last.
In contrast to recursive linearization, our proposed method uses a learned component in place of a key computational bottleneck (namely, applying an expensive preconditioner to the gradient steps; see \cref{subsection:background:classical} for details). As a result, our proposed method runs orders of magnitude faster than recursive linearization while implicitly regularizing the solution to align with prior knowledge of the scattering potential reflected in the collection of training samples.
Empirically, we find that our proposed method yields high-quality reconstructions with errors comparable to, and sometimes lower than, those of more classical approaches but at a fraction of the computational cost;
this represents a significant improvement in accuracy over previous ML methods.

\subsection{Paper outline}
This paper is organized as follows: \Cref{section:background} formally defines the inverse scattering problem and reviews recovery techniques from the literature, focusing on nonlinear least-squares formulations and machine learning methods. \Cref{section:our-method} describes the proposed method, which is empirically analyzed in \cref{section:expts}.

\section{Background}
\label{section:background}
\subsection {Problem setup and notation}
\label{subsection:background:problem}
\begin{figure}
    \centering
    \includegraphics[width=0.4\linewidth]{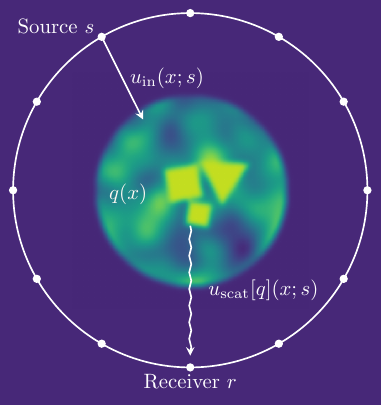}
    \caption{Layout of the wave scattering setup, not to scale: receivers lie much farther from the scattering domain than depicted. An incident plane wave $\uin(x; s)$ from source direction $s$ interacts with the scattering potential $q(x)$, and the resulting scattered wave field $u_{\scat}[q](x; s)$ is recorded by a receiver $r$.
    }
    \label{fig:diagram}
\end{figure}

In inverse medium scattering, 
the goal is to reconstruct an unknown medium based on how it scatters waves (e.g., acoustic or electromagnetic) traveling through it.
A medium's spatially-varying contrast in wave speed is characterized as a scattering potential $q(x)=c_0^2/c^2(x)-1$, where $c_0$ is the wave speed in free space and $c(x)$ is the wave speed in the medium at point $x\in\mathbb R^2$. 
We assume that $q$ is only nonzero within a square domain $\Omega=\left[-\frac{1}{2}, \frac{1}{2}\right]^2$.
The scattering setup is depicted in \cref{fig:diagram}.

The scattering potential $q$ is probed using an incident plane wave $\uin(x;s)=e^{ikx\cdot s}$ with direction $s\in\mathbb S^{1}$, wavelength $\lambda$, and angular spatial frequency $k=2\pi/\lambda$ (also known as the angular wavenumber).
The problem's units are normalized so the wave travels at $c_0\equiv 1$ in free space, and $\lambda=1$ corresponds to the side length of $\Omega$ (equivalently, $\kbar$ is the number of wavelengths per side length of $\Omega$).
The interaction between $q$ and the incident wave $\uin$ results in
an additive perturbation $\usc[q](x;s)$, which is called the scattered wave. The total wave $\utot[q](x;s)=\uin[q](x;s)+\usc[q](x;s)$ is the solution to the following inhomogeneous Helmholtz equation, where the scattered wave satisfies the Sommerfeld radiation boundary condition:

\begin{align}
\left\{
    \begin{aligned}
    \Delta u[q](x; s) + k^2 (1 + q(x)) u[q](x; s) &= 0, && \quad x \in \R^2; \\[6pt]
    \dfrac{\partial \usc[q](x; s)}{\partial \| x \|_2} - i k  \usc[q](x;s) &= o( \| x \|_2^{-1/2}), && \quad \text{as $\|x \|_2 \to \infty$.}
    \end{aligned}
\right.
    \label{eq:pde_problem}
\end{align}

Far-field measurements of the scattered wave are taken at receivers uniformly distributed on a ring with a very large radius ($R \gg 1$) centered at the origin. The measurements take the form
\begin{equation}
d_k(r, s) = (\mathcal F_k[q])(r,s) \equiv \usc[q](Rr; s)
\end{equation}
for each source direction $s\in\mathbb S^1$ and receiver $r\in\mathbb S^1$.
Each scattering potential $q$ is probed with a collection of $N_k$ frequencies $k=k_1,\dots,k_{N_k}$.
It is important to note that the forward model, $\mathcal F_k[q]$, is computationally nontrivial to evaluate, especially when using many source directions.

The objects are represented using the following discretizations, with slight abuse of notation: scattering potential $q\in\mathbb R^{N_x\times N_x}$ has entries $q_{j,\ell}=q(x_j, y_\ell)$, where $x_j, y_\ell$ are each taken from a grid of $N_x$ regular discretization points across each side of $\Omega$.
The array of far-field measurements, $d_k\in\mathbb C^{N_r\times N_s}$, has entries $(d_k)_{j,\ell}=(\mathcal F_k[q])_{j,\ell} =d_k(r_j,s_\ell)$, where the receivers and sources $r_j,s_\ell$ are each taken from a grid of points uniformly spaced about the unit circle $\mathbb S^1$. For simplicity, we take $N_r=N_s$ and use the same set of directions for sources and receivers. In the context of linear operations, we will treat $q$ and $d_k$ as vectors with shapes 
$N_x^2 \times 1$ and $N_rN_s \times 1$, respectively. 
We work in the full-aperture setting, where sources and receivers are distributed across the entire unit circle.

The learning task is to reconstruct a scattering potential given scattered wave measurements taken at a collection of frequencies; i.e., to find a  mapping from $\left( \mathcal F_{k_1}[q], \dots, \mathcal F_{k_{N_k}}[q] \right)$ to $q$. That is, we let $\gennn_\theta$ denote a neural network with learnable parameters $\theta$ and seek to choose $\theta$ so that
\begin{equation}
\hat q_{N_k} = \gennn_\theta\left( \mathcal F_{k_1}[q], \dots, \mathcal F_{k_{N_k}}[q] \right)
\end{equation}
produces good estimates $\hat q$ of $q$.
The network architecture we propose has some learnable components as well as components representing the known forward model, $\mathcal F_{k_t}[\cdot]$, and the adjoint of its Fr\'echet derivative, $\DF_{k_t}[\cdot]^*(\cdot)$; these components are not learned but are instead computed with a numerical PDE solver. We use $(\cdot)^*$ to denote the adjoint.

The training dataset consists of $n$ scattering potentials and their measurements taken at $N_k$ different frequencies:
\begin{align}
\mathcal D
:&= \left\{\left(q^{(j)}, (d_{k_1}^{(j)}, \dots, d_{k_{N_k}}^{(j)}) \right) \right\}_{j=1}^{n} \\
&= \left\{\left(q^{(j)}, (\mathcal F_{k_1}[q^{(j)}], \dots, \mathcal F_{k_{N_k}}[q^{(j)}]) \right) \right\}_{j=1}^{n}.
\end{align}
The wave measurements $(d_{k_1},\dots,d_{k_{N_k}})$ are used as inputs to the neural network, which produces an estimate $\hat q_{N_k}$ of the scattering potential.
This paper uses the same scattering potentials as \cite{melia_multi-frequency_2025}, as these are in a strongly scattering regime, with an inhomogeneous background and a maximum contrast of $\norm{q}_{\infty} = 2$. See \cref{appendix:dataset,appendix:forward-model-equations} for more details.

\subsection{Classical optimization-based approaches}
\label{subsection:background:classical}

\newcommand{\NLLS}{\text{NLLS}}
\newcommand{\FBP}{\text{FBP}}

Because the forward model $\cF_k$ is nonlinear, optimization formulations of the inverse problem involving a least-squares penalty term are in general non-convex. 
In the standard single-frequency nonlinear least squares (NLLS) optimization setting, we seek to solve
\begin{equation}
\argmin_{q} \ell_k(q; d_k)
\quad \text{where} \quad
\ell_k(q; d_k)=\frac {1}{2} \lVert d_k - \mathcal F_k[q] \rVert_2^2.
\label{eq:fwi}
\end{equation}
We let $\hat q_{k}^{\NLLS}$ denote the output of an iterative optimization method applied to this objective.
Iterative algorithms applied to %
the NLLS objective  can provide highly accurate reconstructions, but are often computationally expensive and sensitive to both the algorithm choice and initialization of $q$, making stable recovery in \eqref{eq:fwi} a challenging optimization problem \citep{chauris_seismic_2019,calderon_full-waveform_2022}.
For different incident wave frequencies $k$, the nature of the challenge varies. 
When $k$ is large, the optimization objective becomes highly oscillatory, with many suboptimal local minima, making $\hat q_{k}^{\NLLS}$ highly sensitive to the choice of initialization and resulting in poor reconstructions. 
As $k$ grows, the basin of attraction around the global minimizer of \eqref{eq:fwi} appears to shrink in many low-dimensional numerical examples \citep{bao_numerical_2003,zhou_neural_2023,melia_multi-frequency_2025}.
Simply focusing on small $k$ makes the optimization problem easier but can still yield poor reconstructions because
the corresponding measurements $d_k$ are diffraction-limited and do not capture features in $q$ smaller than a length of about $\lambda/2$ \citep{chen_recursive_1995}.

At low frequencies, the forward model can be approximated as $\mathcal F_k[q]\approx F_k q$, where $F_k\equiv \DF_k[0]$ is the Fr\'echet derivative 
of the forward model centered about $q=0$; \cref{appendix:forward-model-equations} illustrates why this is a reasonable approximation for small $k$.
Under the linear approximation, 
a common approach is to stabilize the inversion using Tikhonov regularization (scaled by a user-selected $\varepsilon > 0$) to yield the filtered back-projection (FBP) method \citep{natterer_mathematics_2001}, which has a closed form solution:
\begin{align}
\hat q_{k}^{\FBP} =& \argmin_q \frac{1}{2}\|d_k - F_k q\|_2^2 + \frac{ \varepsilon}{2} \|q\|_2^2
\\
=&(F_k^* F_k + \varepsilon I)^{-1} F_k^* d_k.
\label{eq:fbp}
\end{align}
As we will discuss in \cref{subsection:background:ml}, the structure of \eqref{eq:fbp} serves as the inspiration for several machine learning methods.
While \eqref{eq:fbp} provides a convenient approximation for low-frequency inversion problems, the estimates $\hat{q}_{k}^{\FBP}$ are often over-smoothed due to the diffraction-limited nature of $\cF_k$.

To alleviate the over-smoothing effects of low-frequency data, a standard strategy is to use data collected at a range of incident wave frequencies \citep{calderon_full-waveform_2022}; we take this approach in the present work. 
Continuation-in-frequency methods order the data from low to high-frequency, solving a sequence of problems like \eqref{eq:fwi} with $k_1 < k_2 < \dots < k_{N_k}$. 
Importantly, the optimization method for the problem at each frequency $k_t$ is initialized using the previous estimate $\hat{q}^{\NLLS}_{k_{t-1}}$. The intuition behind this approach is that when the difference between subsequent frequencies is small, the previous estimate will be near the global optimum of the next problem, so spurious local minima will be avoided.

Recursive linearization is one specific algorithm that follows the strategy of low-to-high frequency reconstructions \citep{chen_recursive_1995,borges_high_2016}. For the sake of clarity, we also present pseudocode in \cref{alg:rl}. The process begins with an initial estimate $\hat q_{k_1}$, for example using filtered back-projection as in \eqref{eq:fbp}, and the update steps for $t=2,3,\dots,N_k$ are given by:
\begin{align}
\delta \hat q_{k_t} 
&=\argmin_{\delta q} \frac {1} {2} \left\lVert d_k - \mathcal F_k[\hat q_{k_{t-1}}]  - \DF_k[\hat q_{k_{t-1}}] \delta q \right\rVert_2^2  \\
&=
\left( \DF_{k_t} [\hat q_{k_{t-1}}]^* \DF_{k_t} [\hat q_{k_{t-1}}] \right)^{-1}
\DF_{k_t} [\hat q_{k_{t-1}}]^*
(d_{k_t} - \mathcal F_{k_t} [\hat q_{k_{t-1}}] )
\label{eq:rl-iter}
\end{align}
where $k_1,k_2,\dots,k_{N_k}$ are increasing frequencies.
Each step folds in information from the frequency band 
$[k_{t-1}, ~k_{t}]$
using the first-order Taylor expansion of the forward model, $\mathcal F_{k_t}[\hat q_{k_{t-1}}] + \DF[\hat q_{k_{t-1}}] \delta q$, centered about the current estimate $\hat q_{k_{t-1}}$. (As a side note: when $\hat q_{k_{t-1}}=0$, the update step reduces to unregularized filtered back-projection, i.e., with $\varepsilon=0$.)
This approximation is referred to as a ``linearization'' of the forward model because it is linear in the perturbation $\delta q$, but it is worth mentioning that the approximation's dependence on $\hat q_{k_{t-1}}$ is in general highly nonlinear (see \cref{appendix:forward-model-equations} for details).
Importantly, the approximation is valid as long as the step $k_{t}-k_{t-1}$ is not too large, even if $k_{t-1}$ and $k_{t}$ themselves are large \citep{chen_recursive_1995}.
Thus recursive linearization enables use of the high-frequency data necessary for high-resolution reconstructions.

In practice, recursive linearization is computationally expensive, since evaluating the forward model $\mathcal F_{k_t}[\cdot]$ involves solving a PDE, and, for a given $\hat q_{k_{t-1}}$, applying the operations $v\mapsto \DF_{k_t}[\hat q_{k_{t-1}}]v$ and $v\mapsto\DF_{k_t}[\hat q_{k_{t-1}}]^*v$ each involve solving an additional PDE for each vector $v$. To invert the large linear system in the filtering step requires many applications of $v\mapsto\DF_{k_t}[\hat q_{k_{t-1}}]v$ and $v\mapsto\DF_{k_t}[\hat q_{k_{t-1}}]^*v$. 
Moreover, it is typical to take small frequency steps: for example, \cite{borges_high_2016} report using steps of $\delta k=0.25$ from $k=1$ to $70$ (corresponding to a maximum non-angular frequency of $\kbar\approx 11$), which requires many CPU core-hours to process a single scattering potential.
Aside from the computational expense, this would also call for the acquisition of data for hundreds of frequencies per sample when using real-world data.

\begin{algorithm}[t]
\SetAlgoLined
\DontPrintSemicolon
\KwData{Measurement data $d_{k_1}, d_{k_2}, \dots, d_{k_{N_k}}$ and regularization parameter $\varepsilon$}
\caption{Recursive linearization algorithm}
\label{alg:rl}
\vspace{8pt}
$\hat q_{k_1} \leftarrow (F_{k_1}^*F_{k_1} + \varepsilon I)^{-1} F_{k_1}^* d_{k_1}$ \tcc*[f]{initialize with FBP}\label{alg:line:fynet}\\
\For{$t=2,3,\dots,N_{k}$}{
    $\gme_{k_t}\leftarrow \DF_{k_t} [\hat q_{k_{t-1}}]^*
(d_{k_t} - \mathcal F_{k_t} [\hat q_{k_{t-1}}] )$  \tcc*[f]{back-projection step}\\
    $\delta \hat q_{k_t} \leftarrow 
    (\DF_{k_t} [\hat q_{k_{t-1}}]^*\DF_{k_t} [\hat q_{k_{t-1}}] )^{-1}
    ~\gme_{k_t}
    $ \tcc*[f]{filtering step}\\
    $\hat q_{k_t} \leftarrow \hat q_{k_{t-1}} + \delta \hat q_{k_t}$
}
\Return{$\hat q_{k_{N_k}}$}
\end{algorithm}

\subsection{Machine learning approaches}
\label{subsection:background:ml}
In recent years, machine learning has been applied effectively to many scientific applications, especially for inverse problems with physics-constrained forward models \citep{arridge_solving_2019,jumper_highly_2021,keith_combining_2021,monga_algorithm_2021,wu_machine_2022,chen_deep_2024}.
In particular, we review several of the machine learning approaches to the inverse scattering problem. Then, we will  discuss the inspiration for our method, which comes from the application of machine learning to linear inverse problems.

For inverse scattering, a common strategy is to design neural network architectures that model the map from measurements to scattering potentials by emulating the structure of filtered back-projection, as presented in \eqref{eq:fbp}. \cite{fan_solving_2022} note that $(F_k^*F_k+\varepsilon I)^{-1}$ is a pseudo-differential operator that can be approximated as a convolution operation and~\cite{zhang_solving_2024} show that the operator is translation-equivariant, motivating the use of a two-dimensional convolutional neural network (CNN).
Examples of architectures inspired by filtered back-projection include 
SwitchNet \citep{khoo_switchnet_2019}, a network we will call FYNet \citep{fan_solving_2022}, WideBNet \citep{li_wide-band_2022}, Wide-band Equivariant Network \citep{zhang_solving_2024}, and (B-)EquiNet-CNN/UNet \citep{zhang_back-projection_2025}.
These methods primarily vary in how they model $F_k^*$ as a neural network, for example by reflecting $F_k^*$'s rotational equivariance or complementary low-rank property.
\citet{zhang_back-projection_2025} also demonstrates that the Wide-band Equivariant architecture can be adapted into a diffusion process as an effective way to sample from the posterior distribution of scattering potentials conditioned on scattering data.
Several other works, such as \citet{sun_efficient_2018} and \citet{khorashadizadeh_deep_2023}, also follow the filtered back-projection structure but avoid emulating $F_k^*$ as a (nonlinear) neural network by taking back-projected measurements $F_k^* d_k$ as inputs rather than $d_k$.
\citet{ong_integral_2022} investigate the application of general-purpose neural operators to inverse scattering tasks in a weakly-scattering regime. 

However, the strategy of emulating filtered back-projection is based on the linear scattering regime; in a more strongly nonlinear scattering regime, it makes sense to consider different approaches.
Loosely inspired by recursive linearization, \citet{melia_multi-frequency_2025} propose a model called MFISNet-Refinement that decomposes the multi-frequency inverse scattering problem into a sequence of steps using data of increasing frequencies to progressively refine estimates of scattering potentials.
Numerical experiments suggest that the progressive refinement structure is beneficial to the model's accuracy.

In this paper, we investigate the explicit use of the forward model alongside neural network components,
as this has greatly improved the reconstruction quality of ML-based methods for linear inverse problems \citep{meinhardt_learning_2017,aggarwal_modl_2019,gilton_neumann_2020,gong_learning_2020}.
For example, \citet{meinhardt_learning_2017} show that a denoising network can be used as a learned prior within the Plug-and-Play priors framework \citep{venkatakrishnan_plug-and-play_2013}.
\citet{gong_learning_2020} propose a deep gradient descent approach for the (linear) deconvolution problem, in which neural networks predict update steps given the gradient of a data-fidelity term, calculated by a differentiable forward model.
Another paradigm is algorithm unrolling \citep{gregor_learning_2010}, in which a sequence of neural network blocks emulate an iterative optimization algorithm, using blocks of fixed weights to implement the gradient of the forward model, and blocks with trainable weights which learn a regularization function from data.
Often, the learned neural network is able to use far fewer layers than the original iterative algorithm would require \citep{monga_algorithm_2021}, which is particularly desirable when the forward model is expensive to evaluate.
Our method incorporates ideas from deep gradient descent and algorithm unrolling in a progressive refinement structure, which helps to avoid the undesirable local minima of the inverse scattering problem.

The inverse scattering literature contains several works employing neural networks along with forward scattering models.
\citet{zhou_neural_2023} use a neural network to warm-start an iterative optimization algorithm with access to the true forward model in an acoustic obstacle scattering problem.
One line of work parameterizes the problem differently, choosing to operate on the quantity $q(x) \cdot u[q](x;s)$, sometimes called the ``contrast source'' or the ``induced current.''
Approaches involving this method include an alternating optimization unrolling \citep{liu_physical_2022}, a subspace-based optimization unrolling \citep{liu_som-net_2022}, and a multi-staged deep learning and iterative optimization approach \citep{sanghvi_embedding_2019}.
In passing we note that \citet{kamilov_plug-and-play_2017} apply the plug-and-play framework to the inverse scattering setting with fixed priors (i.e., no learned components), but the approach may be amenable for use with a deep prior, as \citet{zhang_learning_2017} suggest for the image restoration problem.
In concurrent work, \citet{guo_plug-and-play_2025} pair a deep prior with a differentiable forward model in a Plug-and-Play latent diffusion framework.

Other works avoid calling the expensive forward model, instead opting to approximate or emulate it. \citet{zhou_linear_2020} unroll an alternating optimization algorithm using the (linear) first-order Born approximation to the forward model.
In an unrolled iterative scheme, \citet{guo_physics_2022} learn a forward network for use alongside a neural network that produces updates to the current estimate of the scattering potential.
\citet{zhao_deep_2023} use the Adam optimizer \citep{kingma_adam_2014} to minimize a data-fidelity term where the forward model is replaced with a neural operator as a surrogate.

\section{Our method} \label{section:our-method}

In this section, we provide a formal description of our method, which augments neural network components with the forward model, given by a GPU-accelerated differentiable PDE solver.
Drawing inspiration from the recursive linearization algorithm \citep{chen_recursive_1995}, our method refines a low-frequency initial estimate in a sequence of refinement steps, where each step uses higher-frequency measurement data than the last. This progressive refinement structure helps to avoid falling into low-quality local minima, as discussed in \cref{subsection:background:classical}. Within each refinement step, the forward model computes a local gradient correction to the current estimate (based on a data-fidelity term), and the neural network component filters this to produce an update to the current prediction.

At first glance, our method appears similar to algorithm unrolling~\citep{monga_algorithm_2021}, which alternates between
(i) ``data-fidelity'' steps promoting agreement in the measurement space with explicit use of forward model dynamics, and
(ii) regularization steps implemented with the help of a learnable regularizer, which is trained
in an end-to-end fashion. 
However, the nonlinear-least squares objective \eqref{eq:fwi} is highly nonconvex, meaning that the na{\"i}ve application of algorithm unrolling---namely, using all available measurements in each data fidelity step---will often get stuck with low-quality local minima.

\Cref{fig:block-diagrams} provides a graphical view of our method, illustrating the alternation between learned neural network blocks and fixed numerical PDE solves, while~\cref{alg:our-method} details in pseudocode how the model
makes predictions at inference time. 
We now turn to describing the motivation behind our neural network architecture.
\subsection{Network architecture}
\label{subsection:method:arch}
Our neural network architecture consists of a low-frequency initialization block, followed by a sequence of refinement blocks that alternate between applications of the forward model and a trainable convolutional neural network.
The low-frequency initialization is produced using the FYNet architecture \citep{fan_solving_2022}, though in principle this could be replaced by any neural network designed for the single-frequency setting.

\begin{figure}[t]
    \centering
    \includegraphics[width=\textwidth]{
        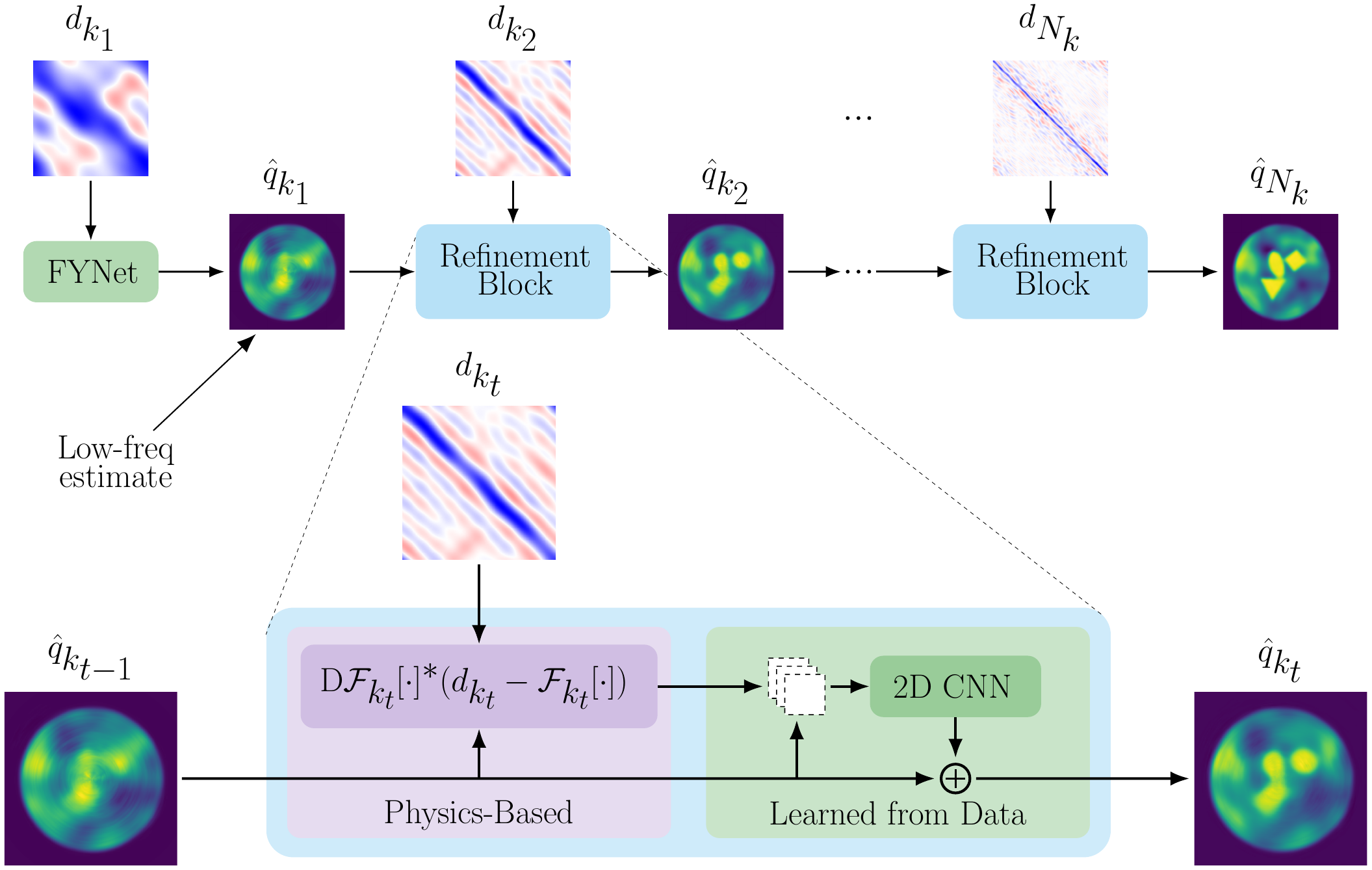
    }
    \caption{Our method's architecture. The top row depicts the overall structure, while the bottom row is a zoomed-in view of each refinement block.
    FYNet is a neural network architecture that takes in single-frequency measurements \citep{fan_solving_2022}.
    }
    \label{fig:block-diagrams}
\end{figure}

Each refinement block learns an additive update to the current estimate of the scattering potential, as reflected in \cref{fig:block-diagrams} by the use of a skip connection.
Within the $t^{\text{th}}$ block, the forward model is used to calculate the negative gradient of the measurement error in~\eqref{eq:fwi}---relative to measurements of frequency $k_{t}$---for the current reconstruction, $\hat{q}_{k_{t-1}}$, which is a steepest descent direction of measurement error:
\begin{equation}
    \gme_{k_{t}}(\hat{q}_{k_{t-1}}; d_{k_{t}}) :=
    -\nabla_{q}\left.\left(\frac{1}{2}
    \norm{d_{k_t} - \mathcal{F}_{k_t}[q]}_2^2
    \right)\right|_{q = \hat{q}_{k_{t-1}}} = \DF_{k_{t}}[\hat{q}_{k_{t-1}}]^* \left(d_{k_t} - \mathcal{F}_{k_t}[\hat{q}_{k_{t-1}}]\right) 
    \label{eq:error-gradient}
\end{equation}
This incurs two calls to the PDE solver: one for $\mathcal F_{k_t}[\hat q_{k_{t-1}}]$, and another to apply $\DF_{k_t}[\hat q_{k_{t-1}}]^* (\cdot)$.
However, directly using $\gme_{k_{t}}$ in a gradient descent scheme results in poor convergence behavior. A standard remedy is to apply a suitable preconditioner---for instance, the recursive linearization algorithm \citep{chen_recursive_1995} uses
$(\DF_{k_t}[\hat{q}_{k_{t-1}}]^* \DF_{k_t}[\hat{q}_{k_{t-1}}])^{-1}$, akin to running one step of a Gauss-Newton method to reduce the squared measurement error \citep{boyd_introduction_2018}; indeed, \cref{alg:rl} can be viewed as a sequence of Gauss-Newton steps on the loss functions $\ell_{k_1}, \dots, \ell_{k_{N_k}}$ defined in~\eqref{eq:fwi}.
However, applying this preconditioner is prohibitively expensive, as every application of the preconditioner requires solving a linear system via an iterative method, wherein \emph{each iteration} involves solving two additional PDEs \citep{borges_high_2016}. %

To bypass this issue, our method uses a neural network to learn a suitable transformation to the gradient vector $\gme_{k_t}(\hat{q}_{k_{t-1}}; d_{k_t})$.
Our architecture uses a convolutional neural network (CNN) following the arguments from~\cite{fan_solving_2022}, who argue that the related preconditioner $(F_{k_t}^* F_{k_t} + \varepsilon I)^{-1}$ is 
amenable to approximation via convolutional neural network blocks.
While we expect $F_{k_t} \neq \DF_{k_t}[\cdot]$ in general, we find that using a cascade of convolutional blocks interspersed with \texttt{ReLU} nonlinearities is sufficient to provide high-quality reconstructions which improve steadily through frequency.
Prior to processing the update vector with the convolutional network, we concatenate $\hat{q}_{k_{t-1}}$ across a new channel dimension to model the dependence of the preconditioner on the current estimate of the scattering potential.

The high-level structure of our method's architecture resembles the MFISNet-Refinement network appearing in~\cite{melia_multi-frequency_2025}, wherein predictions are gradually refined using data of increasing frequencies.
The key difference is that our method uses explicit knowledge of the forward model in the form of the gradient vector, $\gme_{k_t}$, and only learns how to apply a suitable filtering operator (motivated by, but not necessarily equal to, $(\DF_{k_t}[\cdot]^* \DF_{k_t}[\cdot])^{-1}$) from data. Finally,
note that replacing the Gauss-Newton preconditioner by an arbitrary neural network block means that the overall update vector is not automatically guaranteed to be a descent direction for the measurement error; however, as detailed in~\cref{subsection:method:training} below, every update block is trained to approximate the ground truth, making it
unlikely to learn updates that would increase the overall reconstruction error.

\begin{algorithm}[t]
\SetAlgoLined
\DontPrintSemicolon
\KwData{Measurement data $d_{k_1}, d_{k_2}, \dots, d_{k_{N_k}}$ and parameters $\theta_{k_1},\theta_{k_2},\dots,\theta_{k_{N_k}}$}
\caption{
Estimation of one scattering potential from measurements given trained network weights
}
\label{alg:our-method}
\vspace{8pt}
$\hat q_{k_1} \leftarrow \mathtt{FYNet}(d_{k_1}; \theta_{k_1})$ \tcc*[f]{initial estimate}\\
\For{$t=2,3,\dots,N_{k}$}{
    $\gme_{k_t} \leftarrow \DF_{k_t} [\hat q_{k_{t-1}}]^*
(d_{k_t} - \mathcal F_{k_t} [\hat q_{k_{t-1}}] )$ 
    \tcc*[f]{compute gradient with PDE solver}\\
    $\delta \hat q_{k_t} \leftarrow \mathtt{CNN}(\hat q_{k_{t-1}}, \gme_{k_t}; \theta_{k_t})$ \label{alg:line:nn} \tcc*[f]{apply learned filter} \\
    $\hat q_{k_t} \leftarrow \hat q_{k_{t-1}} + \delta \hat q_{k_t}$
}
\Return{$\hat q_{k_{N_k}}$}
\end{algorithm}

\subsection{Forward model} \label{subsection:method:forward-model}
In this section, we discuss computation of the forward model, $\mathcal{F}_{k}[\cdot]$, as well as vector-Jacobian products of the form $v \mapsto \DF_{k}[q]^* v$, where $q$ is an arbitrary scattering potential.
To evaluate these quantities, we use a GPU-accelerated implementation of the Hierarchical Poincar{\'e}-Steklov (HPS) solver from~\cite{melia_hardware_2025}, which is made available via the \texttt{jaxhps} package \citep{melia_jaxhps_2025}.
The HPS solver is direct method which discretizes the interior problem in \eqref{eq:pde_problem} using a composite spectral collocation scheme and enforces the Sommerfeld radiation condition from \eqref{eq:pde_problem}
with a boundary integral equation \citep{gillman_spectrally_2015}.
To evaluate the forward model $q \mapsto \mathcal{F}_{k}[q]$, the HPS solver computes and stores a factorization of the differential operator in \eqref{eq:pde_problem}. This differential operator is self-adjoint, so the factorization can be re-used to compute vector-Jacobian products $v \mapsto \DF_{k}[q]^* v$ with a small amount of extra work~\citep{borges_high_2016}. Because of this fact, we elide the reference to the derivative and simply refer to our calls to HPS as calls to the forward model when there is no ambiguity.
We provide additional details on the defining equations of the forward model in~\cref{appendix:forward-model-equations}.

\subsection{Training strategy} \label{subsection:method:training}

Our method iterates over frequencies $k_t$ for $t = 1, \dots, N_{k}$, adjusting the weights of each refinement block in order of appearance. 
In particular, the $t^{\text{th}}$ stage only involves the training and evaluation of the $t^{\text{th}}$ refinement block, and all other refinement blocks are unused.
The block's parameters, $\theta_{k_t}$, are adjusted by minimizing the average squared reconstruction error over the training set:
\begin{equation}
    \mathcal{L}_{k_{t}}(\theta_{k_t}) = \begin{cases}
    \displaystyle
        \frac{1}{n} \sum_{j = 1}^{n}
        \norm[\big]{q^{(j)} - \mathtt{FYNet}(d_{k_t}^{(j)}; \theta_{k_t})}^2,
        & \text{for $t = 1$}; \\[1em]
    \displaystyle
        \frac{1}{n} \sum_{j = 1}^{n}
        \norm[\big]{q^{(j)} - \mathtt{CNN}(\hat{q}_{k_{t-1}}^{(j)}, \gme_{k_{t}}({\hat{q}_{k_{t-1}}}^{(j)}; d_{k_{t}}^{(j)}); \theta_{k_t})}^2,
        & \text{for $t = 2, 3, \dots, N_{k}$.}
    \end{cases}
    \label{eq:cost-functions}
\end{equation}
Any iterative method may be used to minimize the cost functions in~\eqref{eq:cost-functions}. We use the \texttt{AdamW} optimizer \citep{loschilov_decoupled_2019} with suitably chosen hyperparameters; for more details, we refer the reader to \cref{section:expts} for the experimental setup and \cref{appendix:nn-hyperparams} for hyperparameter values. Finally, we note in passing that prior work~\citep{melia_multi-frequency_2025} considered replacing the targets $q^{(j)}$ used in~\eqref{eq:cost-functions}
with progressively finer approximations of the true scattering potentials,
in a bid to achieve more stable reconstructions. However, we find that this strategy
confers no benefits to our training method.

As outlined in \cref{alg:training-blockwise}, 
we train the neural network in a block-wise manner, which sidesteps the need to backpropagate gradients through the HPS solver. Backpropagating through the HPS evaluations of the forward model $q \mapsto \mathcal{F}_k[q]$ and vector-Jacobian products $x \mapsto \DF_{k}[q]^* x$
every training step
would incur a large memory footprint in the multi-frequency setting, which is the focus of the present paper.

\begin{algorithm}[t]
\SetAlgoLined
\DontPrintSemicolon
\KwData{Dataset $\mathcal D = \left\{\left(q^{(j)}, ~~(d_{k_1}^{(j)}, d_{k_2}^{(j)}, \dots, d_{k_{N_k}}^{(j)}) \right)\right\}_{j=1}^n $}
\caption{Block-wise training}
\label{alg:training-blockwise}
\vspace{4pt}
Randomly initialize parameters $\theta_{k_1},\theta_{k_2},\dots,\theta_{k_{N_k}}$ \\
Train $\mathtt{FYNet}(\cdot; \theta_{k_1})$ on dataset $\mathcal D_{k_1} = \{(q^{(j)}, ~d_{k_1})\}_{j=1}^n$ with loss \eqref{eq:cost-functions} \\
$\hat q_{k_{1}}^{(j)} \leftarrow \mathtt{FYNet}(d_{k_1}^{(j)}; \theta_{k_1})$ for $j=1,\dots,n$ \tcc*[f]{Low-frequency predictions} \\
\For{$t=2,3,\dots,N_{k}$}{
    Compute $\gme_{k_t}^{(j)} = \gme_{k_t}(\hat q_{k_{t-1}}^{(j)}; d_{k_{t}}^{(j)})$ for $j=1,\dots,n$ with \eqref{eq:error-gradient} and PDE solver \\
    Train $\mathtt{CNN}(\cdot; \theta_{k_t})$ on dataset $\mathcal D_{k_t} = 
    \left\{\left(q^{(j)}, ~~(\hat q_{k_{t-1}}^{(j)}, \gme_{k_t}^{(j)})\right)
    \right\}_{j=1}^n$ with loss \eqref{eq:cost-functions} \\
    $\delta \hat q_{k_{t}}^{(j)} \leftarrow \mathtt{CNN}(\hat q_{k_{t-1}}^{(j)}, \gme_{k_t}^{(j)}; \theta_{k_t})$ for $j=1,\dots,n$ \\
    $\hat q_{k_{t}}^{(j)} \leftarrow \hat q_{k_{t-1}}^{(j)} + \delta \hat q_{k_{t}}^{(j)}$ for $j=1,\dots,n$
    \tcc*[f]{Update predictions}
}
\Return{$\hat q_{k_{N_k}}$}
\end{algorithm}

\section{Numerical experiments}
\label{section:expts}

In this section, we present several numerical experiments to help characterize the behavior of the proposed method, dubbed {\hpscnn} for its joint use of the HPS solver as the forward model and 2D CNNs as the learned components.
We work with the dataset of scattering potentials from \cite{melia_multi-frequency_2025}.
These potentials have high contrast and an inhomogeneous background, making the inverse scattering problem particularly challenging;
see \cref{appendix:dataset} for more details and \cref{fig:sample-scobjs} for a few examples.
Each potential is observed at $N_k=10$ frequencies, for $\kbar=1,2,\dots,10$. For most experiments (with the exception of~\cref{subsection:expts:sample-complexity}), we use a dataset with $N_{\mathrm{train}} = 1000$ training samples. Additionally, we have 
a validation set comprising $N_{\mathrm{val}}=1000$ samples to select hyperparameters such as learning rate, and $N_{\mathrm{test}}=1000$ test samples to evaluate the final model performance.

We hypothesize that the incorporation of a differentiable PDE solver, reflecting precise knowledge of the dynamics of the scattering process, confers multiple advantages. In particular:
\begin{itemize}
\item We expect to outperform ML-based baselines that do not use knowledge of the problem physics and to compete with or surpass (in terms of reconstruction quality) classical methods based on nonlinear least-squares, whose complexity at test time is significantly higher; see~\cref{subsection:expts:baselines} for an overview of baseline methods and an initial comparison.
\item We posit that the \emph{sample efficiency} of the new method is improved relative to other ML baselines: indeed, {\hpscnn} need only learn a nonlinear correction to descent directions for the squared measurement errors, while competing methods ``learn'' the entire reconstruction pipeline in an end-to-end manner; see~\cref{subsection:expts:sample-complexity}.
\item We anticipate that knowledge of the forward model endows the overall method with greater stability to misspecification. We test this hypothesis in two scenarios: namely, under additive measurement noise (see~\cref{subsection:expts:noisy}), and under distribution shift at test time (see~\cref{subsection:expts:ood}).

\item Finally, we expect that the progressive refinement strategy adopted by our method is essential to avoid local minima and to furnish high-quality reconstructions.
We verify this numerically in~\cref{subsection:expts:prog-ref-abl}, where we effectively disable progressive refinement by forcing each update block to access measurement data at the highest available frequency.
\end{itemize}

\begin{figure}
\centering
\includegraphics[width=0.8\textwidth]{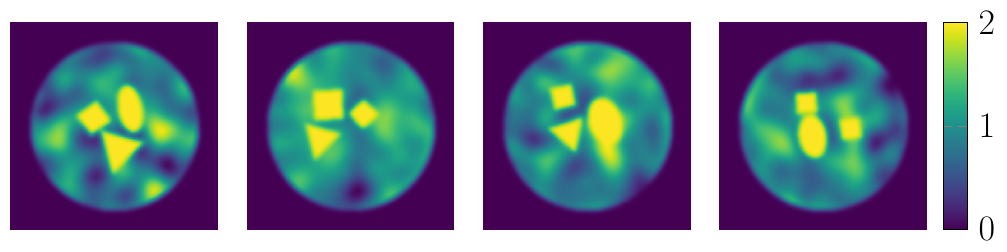}
\caption{Several example scattering potentials from the dataset.}
\label{fig:sample-scobjs}
\end{figure}

\subsection{Baseline methods}
\label{subsection:expts:baselines}
In this section, we discuss the reconstruction methods that we use as baselines for our numerical experiments. Two of these methods are ML-based: namely, the MFISNet-Refinement neural network from~\citep{melia_multi-frequency_2025}, which we will shorten to MFISNet in figures, as well as the uncompressed variant of the Wide-band Equivariant Network from~\citep{zhang_solving_2024}, which we shorten to WBENet in figures.
We use the implementation of the Wide-band Equivariant Network from \citet{zhang_baseline_2025}, with adaptations to our problem setting as described in \cref{appendix:nn-hyperparams}.
The former is a neural network designed to emulate the progressive refinement structure of the recursive linearization method, albeit without any explicit knowledge of the forward model; 
the latter back-projects multi-frequency measurement data using a learned kernel, enabling a parameter-efficient mapping to scattering potentials.
We use two variants of MFISNet-Refinement, which differ in their use of low-pass filtered or unmodified targets to guide the intermediate reconstructions of the network.
The original version of MFISNet-Refinement was trained against smoothed intermediate targets \citep{melia_multi-frequency_2025}, but we include the variant trained on unsmoothed intermediate targets to ensure that {\hpscnn}'s advantage is not solely due to its use of unsmoothed intermediate targets.
We note that, like the {\hpscnn} method, all the above baselines use a 2D CNN to approximate the filtering operations in~\eqref{eq:rl-iter}; our experiment can thus help isolate the effect of 
explicitly using the forward model.

In addition, we compare our method to a classical method based on nonlinear least-squares. %
First, we compare against the recursive linearization method outlined in~\citep{borges_high_2016}, 
although
we represent all our scattering potentials in a $N_{x} \times N_{x}$ pixel basis, instead of the frequency-dependent 2D sine basis employed in~\citep{borges_high_2016}, and employ Tikhonov regularization to compensate for the increased basis size. We call this ``Original RecLin''.
Second, we consider a variant that performs multiple Gauss-Newton steps per frequency as well as Tikhonov regularization (as a side note, Gauss-Newton with Tikhonov regularization is equivalent to the Levenberg-Marquardt algorithm \citep{boyd_introduction_2018}).
We call this ``Modified RecLin''. These modifications are necessary to achieve good performance in our setting because we only use ten frequencies, while
the original recursive linearization method is typically employed in settings with
very small frequency spacing between available frequencies and consequently more frequencies overall.
Note that Modified RecLin is not a method appearing elsewhere in the literature, but rather a method we developed as a baseline for this paper to help us better quantify the impact of training data on performance.
See~\cref{appendix:rl-impl} for more details, including hyperparameter tuning.

\begin{figure}[t!]
    \centering
    \includegraphics[width=0.8\textwidth]{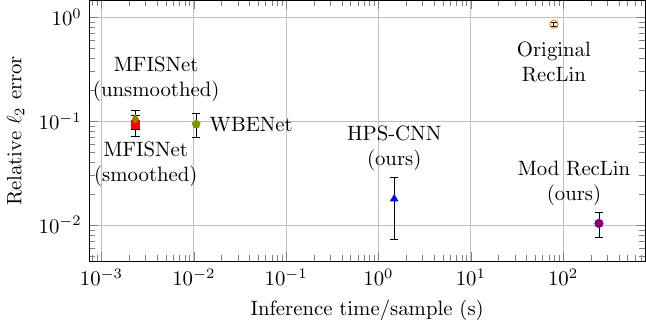}
    \caption{Relative test error vs. inference time per sample, averaged over 1000 samples.
    Timing performed on an NVIDIA\textsuperscript{\textregistered} RTX A6000 GPU.
    }
    \label{fig:err-vs-time}
\end{figure}

The results of our experiment, following the setup described in the beginning of~\cref{section:expts}, are illustrated in~\cref{fig:err-vs-time} along with inference times.
\cref{appendix:additional-results} contains additional details in \cref{tab:basic-expt,tab:train-inference-times}, as well as sample visualizations of predicted scattering potentials \cref{fig:pipeline-preds-1k,fig:pipeline-preds-10k}.
When comparing with ML baselines, we find that the {\hpscnn} method achieves moderately lower train error and significantly lower test error,
by a factor of about five.
As can be seen in \cref{tab:basic-expt} in the appendix, the gap between train and test error is significantly smaller for {\hpscnn}, suggesting that the new method is less prone to overfitting. In light of this, we anticipate it to perform favorably in our extended experiments in subsequent sections. 

The comparison with our recursive linearization implementation paints a more subtle picture. On one hand, we see that the original version of recursive linearization is unable to achieve nontrivial test error, which is unsurprising given the moderate spacing between frequencies in our experiments. On the other hand, the modified version, which employs multiple Gauss-Newton steps per frequency as well as Tikhonov regularization, achieves smaller test error relative to {\hpscnn}; however, the time spent on reconstructing each sample at test time is significantly larger, as suggested by~\cref{fig:err-vs-time} and \cref{tab:train-inference-times}. In particular, {\hpscnn} takes less than 2 seconds to reconstruct a single potential, while modified recursive linearization takes 4 minutes on average, two orders of magnitude longer.
We also note that while the training stage of {\hpscnn} takes about 200 minutes, this can be amortized over a set of test samples, whereas this is not possible for recursive linearization algorithms.
This also means that, including training, {\hpscnn} is faster than Modified RecLin when making predictions on more than about 50 new samples.

\subsection{Sample efficiency}
\label{subsection:expts:sample-complexity}
In our next experiment, we study the sample efficiency of various reconstruction methods. In particular, we vary the size of the training set $N_{\mathrm{train}}$ from $10^{2}$ to $10^{4}$ scattering potentials and compute the average relative $\ell_{2}$
reconstruction error on a held-out test set which is held fixed throughout for each value of $N_{\mathrm{train}}$.

\begin{figure}[t!]
    \centering
    \includegraphics[width=0.8\textwidth]{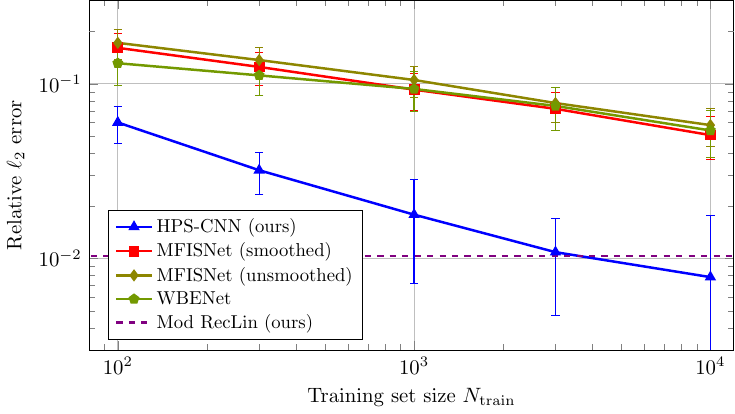}
    \caption{Sample efficiency illustrated by relative test error as a function of training set size.
    }
    \label{fig:sample-complexity}
\end{figure}

The results of the experiment are illustrated in~\cref{fig:sample-complexity} (also see~\cref{tab:sample-complexity}), wherein it is apparent that the proposed method consistently outperforms ML baselines across training set sizes.
Also note that to achieve a desired level of accuracy, our method requires significantly less training data than the competing ML baselines.
At the same time, it exhibits favorable scaling with $N_{\mathrm{train}}$, with the gap between itself and competing methods widening as the training set size increases, and even outperforming the non-ML baseline for $N_{\mathrm{train}} = 10^{4}$.

\subsection{Robustness to noise}
\label{subsection:expts:noisy}

Our next experiment examines the effect of measurement noise on the reconstruction error. In particular, we assume that each measurement is corrupted by additive Gaussian noise at some specified level $\sigma$, resulting in a set of observed measurements $\bar{d}_k \in \mathbb{C}^{N_r \times N_s}$ satisfying
\begin{equation}
    \bar{d}_{k} = d_{k} 
    + \sigma \norm{d_k}_\infty
    \cdot \frac{Z_1 + i Z_2}{\sqrt{2 N_r N_s}},
    \quad \text{where} \quad [Z_{j}]_{r,s} \overset{\mathrm{i.i.d.}}{\sim} \mathcal{N}(0, 1),
    \label{eq:noise-model}
\end{equation}
The noise model in~\eqref{eq:noise-model} satisfies
$\mathbb{E}_{Z}\left[\norm{\bar{d}_k - d_k}_2 / \norm{d_k}_\infty\right] \approx \sigma$, corresponding to a PSNR of $-20 \log_{10} (\sigma)$ decibels. 
For this experiment only, we train and test our models using noisy versions of the dataset. Note that while the measurements are noisy, the scattering potential targets are not corrupted by noise.

\begin{figure}[t]
    \centering
    \includegraphics[width=0.8\textwidth]{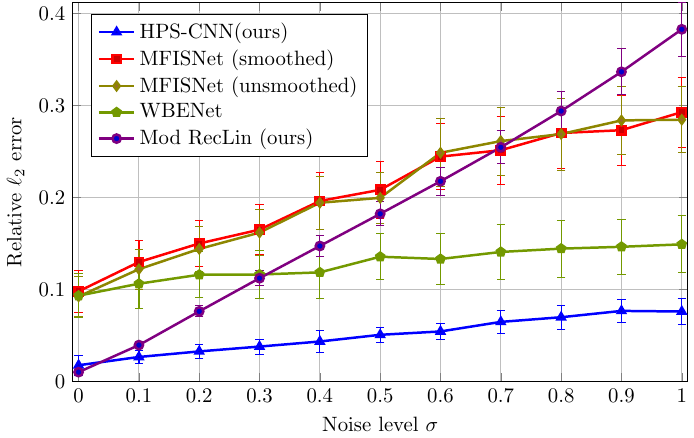}
    \caption{Performance under measurement noise given varying noise ratio $\sigma$.
    }
    \label{fig:noise-robustness}
\end{figure}

Our experiments, illustrated in~\cref{fig:noise-robustness} (see also~\cref{tab:noisy}), suggest that the performance of {\hpscnn} scales with $\sigma$ at a rate that is on par with, if not better than, other competing ML methods, and much more slowly than Modified RecLin. In particular, all ML baselines are able to use the information in the training set to learn a mapping from noisy measurements to relatively noise-free reconstructions, with {\hpscnn} consistently achieving the lower reconstruction error.
This experiment illustrates that the learned components of {\hpscnn} not only improve reconstruction speed, but also greatly improve robustness to noise.
We do not depict the performance of Original RecLin in the plot, since it achieves a relative $\ell_2$ 
error of more than $0.8$ at every noise level.

\subsection{Robustness to distribution shift}
\label{subsection:expts:ood}
In this experiment,
we probe the resilience of our reconstruction method to distribution shift at test time. Starting from the test set used in~\cref{subsection:expts:baselines}, we create a collection of test sets that are identical in all aspects but one: namely, the contrast level $\norm{q}_{\infty}$. We train all ML methods (and select hyperparameters for the recursive linearization methods) on a dataset with contrast $\norm{q}_{\infty} = 2$ and test all methods
on a variety of contrast levels between $0.5$ and $3.0$.
Note that a small change in the contrast can yield large changes in the measurements because of the nonlinearity of the forward model, and this also affects the conditioning of the reconstruction problem. Thus, this form of distribution drift can be challenging for ML-based approaches.

\begin{figure}[t!]
    \centering
    \includegraphics[width=0.8\textwidth]{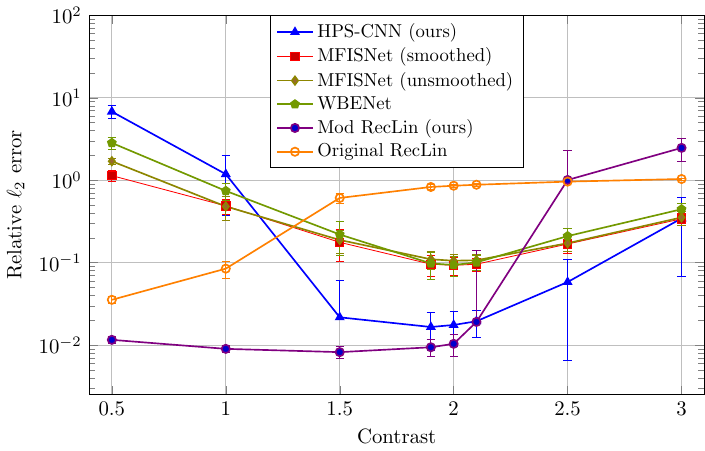}
    \caption{Test-time performance on samples with varying contrast levels $\norm{q}_{\infty}$. All samples in the training set have contrast $\norm{q}_{\infty} = 2$.}
    \label{fig:distribution-shift}
\end{figure}

The results are illustrated in~\cref{fig:distribution-shift}. 
All ML methods exhibit nontrivial sensitivity to distribution shift, with {\hpscnn} emerging as the most stable for contrast levels near that of the training set, with performance even seeing a slight improvement at contrast 1.9 compared to the original 2.0. In addition, the MFISNet model trained on low-pass filtered intermediate targets appears to be the most resilient to extreme changes in contrast level, especially when the target contrast is significantly smaller than the one used to train the model.
The Modified RecLin method, which does not use machine learning (but does involve several hyperparameters selected at contrast 2), performs well up to about contrast 2, but then the performance quickly degrades.

\subsection{Progressive refinement ablation}
\label{subsection:expts:prog-ref-abl}
Finally, we present a small ablation experiment, where we remove the progressive refinement structure from our method to probe its importance to the method.
Instead of the standard input data of $\kbar=1,2,\dots,10$, we use measurements collected at $\kbar=10$ in \emph{every} refinement block.
In each case the same architecture is used, with ten neural network blocks: one FYNet block followed by nine refinement blocks.
\cref{tab:prog-ref-abl} reveals that removing the progressive refinement structure severely impacts the reconstruction error of the method, thus illustrating its importance to our method. 

\begin{table}[t!]
    \centering
    \begin{tabular}{lLRR}
        \toprule
            \makecell{\textbf{Progressive}\\\textbf{refinement?}} 
            & \textbf{Input frequencies}
            & \textbf{Test error} \\
        \midrule
            Yes (standard) & \kbar=1,2,\dots,10
            & 1.783\% \pm 1.1\% \\
            No & \kbar=10,10,\dots,10
            & 28.082\% \pm 9.5\% \\
        \bottomrule \\
    \end{tabular}
    \caption{
        Ablation experiment removing the progressive refinement structure from our method.
        Both settings use the same architecture and choice of hyperparameters.
    }
    \label{tab:prog-ref-abl}
\end{table}

\section{Conclusion}
\label{section:conclusion}

In this article, we present a machine learning method for the inverse scattering problem that produces reconstructions with quality competitive with classical baselines that require orders of magnitude more computation time.
The method succeeds due to its combined use of physical knowledge, through evaluations of the forward model and progressive refinement structure, and implicit data-driven regularization of scattering potentials, using learned components.
Additionally, despite the relatively high computational expense of calls to the forward model, our block-wise training strategy allows for a much faster overall training stage than would be expected from an end-to-end training strategy.

Our work leaves open a number of directions for future investigation.
Our architecture is inspired by the recursive linearization algorithm, but there may be alternative architectures that draw on other ideas from classical nonlinear image reconstruction approaches in order to improve the conditioning of the problem.
Additionally, adapting the method to different problem settings may extend its applicability: for example, measurement data may have limited aperture, lack phase information, or be taken in 3D.

\section*{Acknowledgements}
OM, OT, VC, and RW gratefully acknowledge the support of AFOSR
FA9550-18-1-0166, NSF DMS-2023109, 
and
the NSF-Simons AI-Institute for the Sky (SkAI) via grants NSF AST-2421845 and Simons Foundation MPS-AI-00010513. RW and YK gratefully acknowledge the support of
DOE DE-SC0022232. YK gratefully acknowledges the support of NSF DMS-2339439. 
JH was supported in part by a Sloan Research Fellowship.
The team gratefully acknowledges the support of the Margot and Tom Pritzker Foundation.
The Flatiron Institue is a division of the Simons Foundation. 

\bibliographystyle{elsarticle-harv} 
\bibliography{reference}

\appendix
\section*{Appendices}
\addcontentsline{toc}{section}{Appendices}
\renewcommand{\thesubsection}{\Alph{subsection}}

\subsection{Dataset}
\label{appendix:dataset}
The dataset shares scattering potentials with \citet{melia_multi-frequency_2025}. The same maximum contrast level of $\norm{q}_\infty=2$ is used everywhere, except for the distribution shift experiment in \cref{subsection:expts:ood}. For reference, this is equivalent to an index of refraction of $c_0/c=\sqrt{2+1}\approx 1.73$.

Measurements are generated with a solver based on the Lippmann-Schwinger integral equation, as in \cref{eq:ls-near,eq:ls-far,eq:ls-fwd-model} using a custom PyTorch \citep{ansel_pytorch_2024} implementation of BiCGSTAB \citep{van-der-vorst_bi-cgstab_1992} for the GPU, and the linear system is solved to a relative tolerance of $10^{-4}$ for each source direction. The use of this iterative solver enables finer control over solution accuracy as compared with the direct HPS solver used by our proposed method. The train, validation, and test sets contain $N_\mathrm{train}=10,000$, $N_\mathrm{val}=1,000$, and $N_\mathrm{test}=1,000$ scattering potentials, respectively. Measurements are taken at with angular spatial frequencies $\kbar=1,2,\dots,10$, which correspond to waves with $1,2,\dots,10$ wavelengths across the scattering domain $\Omega$.
Note that these frequencies differ from the frequencies used by \citet{melia_multi-frequency_2025}, which are $\kbar=1,2,4,8,16$.

FYNet \citep{fan_solving_2022} and MFISNet-Refinement \citep{melia_multi-frequency_2025} represent the scattering potentials on a polar grid of size $N_\rho \times N_\theta = 96 \times 192$, and we observed that at low frequencies (like $\kbar=1,2$), the low-pass filtered targets extend outside the original polar grid whose maximum radius was $\rho_{max}=0.5$. As a fix, we stretch the polar grid to have a maximum radius of $\rho_{max}=0.575$ while maintaining the number of grid points. The dataset also contains the same scattering potentials on a Cartesian grid, which is used by our proposed method, Wide-band Equivariant Network \citep{zhang_solving_2024}, and both versions of recursive linearization.

Scattered wave measurements $d_k$ are similarly represented in two different coordinate systems. The first is a grid with dimensions $N_r \times N_s=192 \times 192$, as described in \cref{subsection:background:problem}, and is used by our method's refinement blocks, Wide-band Equivariant Network, and the recursive linearization methods. The second system, used by MFISNet-Refinement \citep{melia_multi-frequency_2025} and our method via FYNet \citep{fan_solving_2022}, describes the data with re-parameterized coordinates $m=(r+s)/2$ and $h=(r-s)/2$ on a grid of size $N_m \times N_h=192 \times 96$.

\subsection{Additional result tables and figures}
\label{appendix:additional-results}
\begin{sidewaysfigure}[h!]
    \centering
    \includegraphics[width=\linewidth]{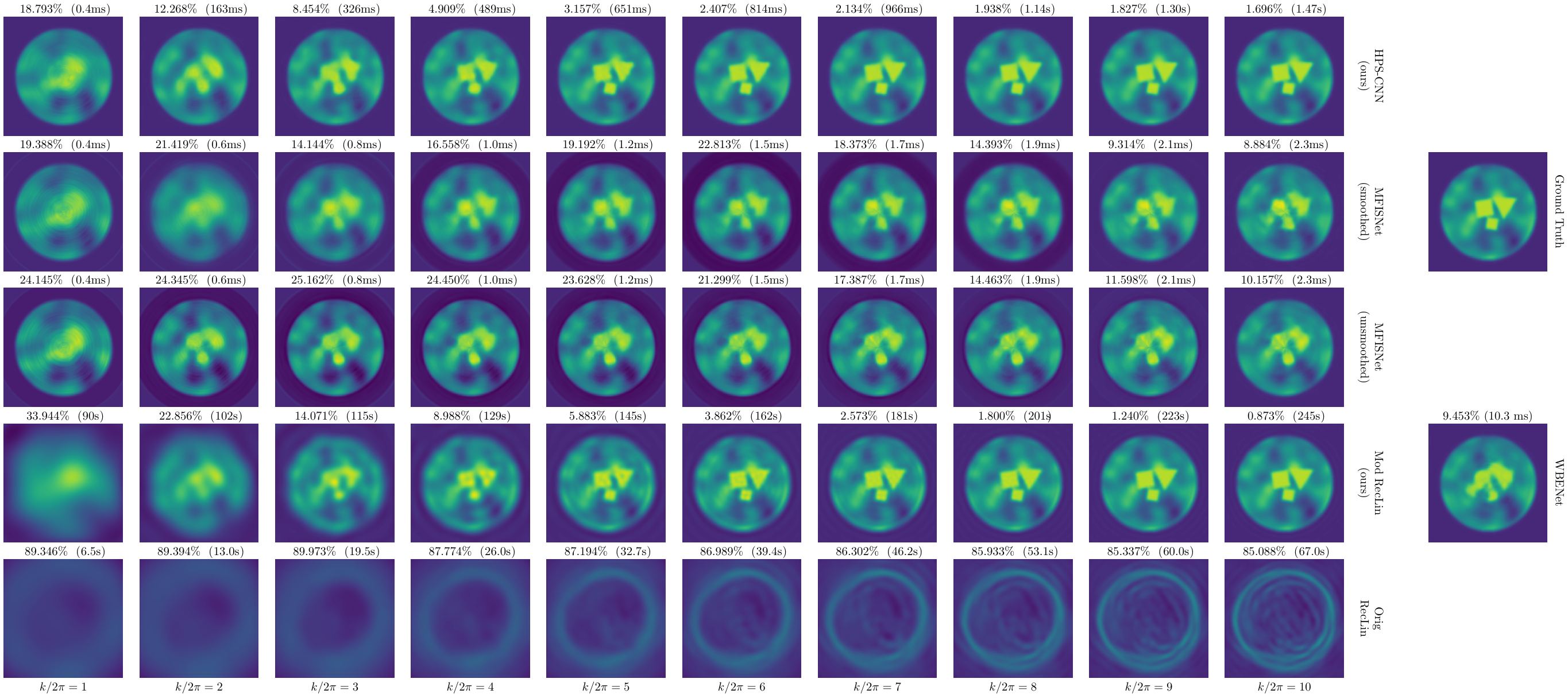}
    \caption{Sample predictions from models trained on $N_{\mathrm{train}}=1000$ samples, including intermediate predictions where relevant.}
    \label{fig:pipeline-preds-1k}
\end{sidewaysfigure}

\begin{sidewaysfigure}[h!]
    \centering
    \includegraphics[width=\linewidth]{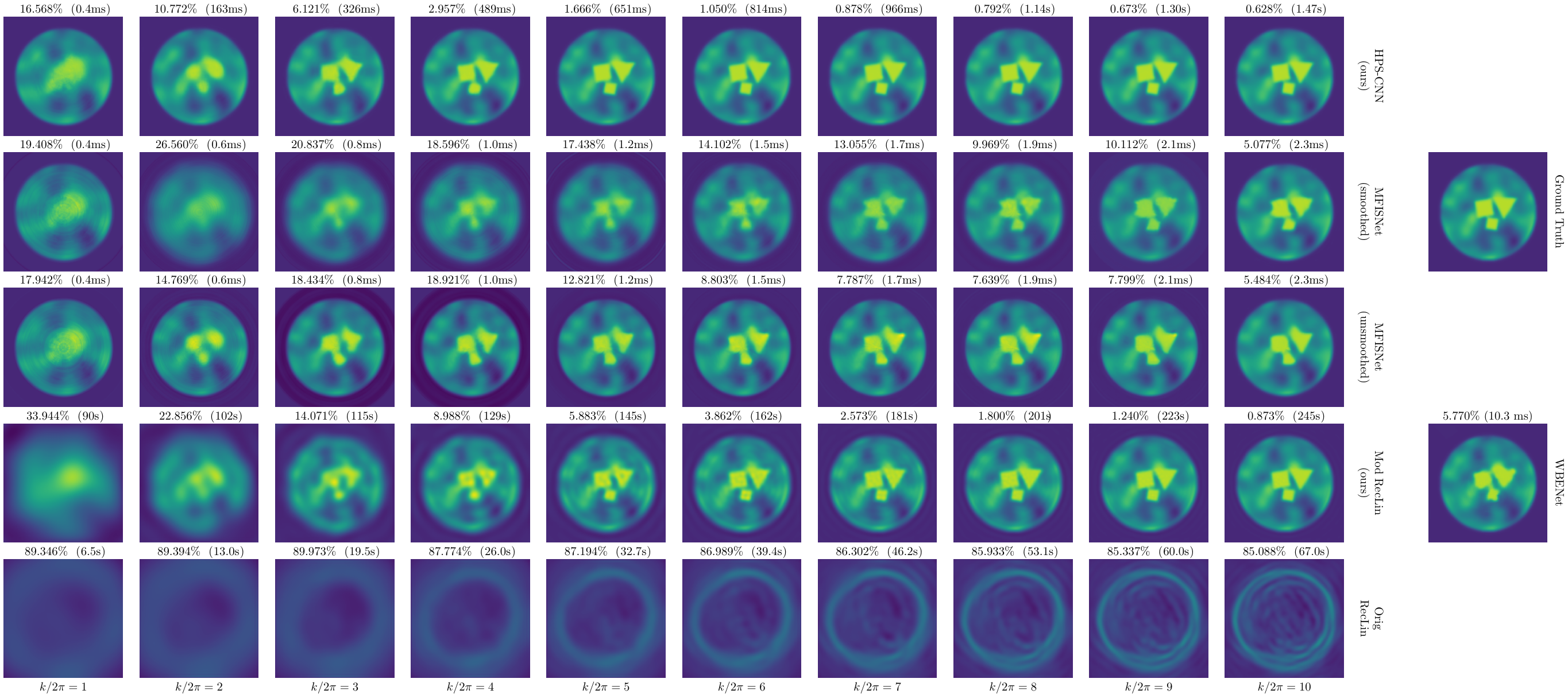}
    \caption{Sample predictions from models trained on $N_{\mathrm{train}}=10000$ samples, including intermediate predictions where relevant.}
    \label{fig:pipeline-preds-10k}
\end{sidewaysfigure}

We provide the experimental results in table format as~\cref{tab:basic-expt,tab:train-inference-times,tab:sample-complexity,tab:noisy,tab:ood}. Additionally, we include intermediate predictions for a test sample in \cref{fig:pipeline-preds-1k} and \cref{fig:pipeline-preds-10k}, where models are trained on $N_\mathrm{train}=1000$ and $10000$ samples, respectively.

\begin{table}[t]
    \centering
    \begin{tabular}{lrr}
        \toprule
            \textbf{Method Name} & \textbf{Train error} & \textbf{Test error} \\
        \midrule
            {\mmgunet} (ours)      & $1.662\% \pm 0.4\%$ & $1.783\% \pm 1.1\%$ \\
            {MFISNet} (smoothed)   & $3.459\% \pm 0.3\%$ & $9.258\% \pm 2.2\%$ \\
            {MFISNet} (unsmoothed) & $4.440\% \pm 0.5\%$ & $10.516\% \pm 2.1\%$ \\
            {WBENet}               & $7.661\% \pm 1.3\%$ & $9.345\% \pm 2.4\%$ \\
            Modified RecLin (ours) & - & $1.035\% \pm 0.3\% $\\
            Original RecLin & - & $86.082\% \pm 3.9\%$ \\
         \bottomrule \\
    \end{tabular}
    \caption{
    Comparison of different methods by relative $\ell_2$ error of predictions, presented as the mean and standard deviation over train and test sets.
    The neural networks here are trained with $N_\textrm{train}=1000$.
    }
    \label{tab:basic-expt}
\end{table}

\begin{table}[h]
    \centering
    \caption*{\textbf{Train and inference times}}
    \resizebox{\textwidth}{!}{%
    \begin{tabular}{lrr}
        \toprule
            \textbf{Method name}
        & Train time ($N_\textrm{train}=1000$)
            & Inference time/sample ($N_\textrm{test}=1000$)\\
        \midrule
        {\mmgunet} (Ours)      & 200m & 1.48s \\
        MFISNet (smoothed)     &  80m & 2.3ms \\
        MFISNet (unsmoothed)   &  80m & 2.3ms \\
        WBENet                 &  70m & 10.3ms \\
        Original RecLin        &    - & 67s \\
        Modified RecLin (Ours) &    - & 245s \\
        \bottomrule
    \end{tabular}}
    \vspace{0.5em} 
    \caption{Time required for training and inference per sample. Training time includes the evaluation of the validation set, as it determines which epoch to select for model weights. Our method spends roughly 25\% of the time running HPS and the rest training the neural network components.}
    \label{tab:train-inference-times}
\end{table}

\begin{table}[h]
    \centering
    \caption*{\textbf{Sample efficiency}}
	\resizebox{1.0\textwidth}{!}{
    \begin{tabular}{lRRRRR}
        \toprule
            \textbf{Method name}
            & N_{\mathrm{train}} = 100
            & N_{\mathrm{train}} = 300
            & N_{\mathrm{train}} = 1000
            & N_{\mathrm{train}} = 3000 &
            N_{\mathrm{train}} = 10000 \\
        \midrule
        {\mmgunet}(Ours)
        &  6.009\% \pm 1.4\% &  3.202\% \pm 0.9\% &  1.783\% \pm 1.1\% & 1.088\% \pm 0.6\% & 0.783\% \pm 1.0\% \\
        MFISNet (smoothed)
        & 16.062\% \pm 3.3\% & 12.490\% \pm 2.7\% &  9.258\% \pm 2.2\% & 7.191\% \pm 1.8\% & 5.087\% \pm 1.4\% \\
        MFISNet (unsmoothed)
        & 17.154\% \pm 3.4\% & 13.672\% \pm 2.5\% & 10.516\% \pm 2.1\% & 7.760\% \pm 1.7\% & 5.802\% \pm 1.4\% \\
        WBENet
        & 13.116\% \pm 3.3\% & 11.173\% \pm 2.6\% & 9.345\% \pm 2.4\% & 7.471\% \pm 2.1\% & 5.407\% \pm 1.6\% \\
        \bottomrule
    \end{tabular}
	}
    \vspace{0.5em}
    \caption{Comparison of different methods trained on varying training set sizes, all in a noiseless setting. {\mmgunet} consistently achieves significantly lower errors than the other ML baselines.
    }
    \label{tab:sample-complexity}
\end{table}

\begin{table}[h]
    \centering
    \caption*{ \textbf{Performance with noise}}
	\resizebox{\textwidth}{!}{
    \begin{tabular}{lRRRRRR}
        \toprule
            \text{Noise level}
            & \makecell{ \text{{\hpscnn}} \\ \text{(ours)}  }
            & \makecell{\text{MFISNet} \\ \text{(smoothed)}}
            & \makecell{\text{MFISNet} \\ \text{(unsmoothed)}}
            & \text{WBENet}
            & \makecell{ \text{Mod RecLin} \\ \text{(ours)} }
            & \text{Orig RecLin} \\
        \midrule
            0.0 & 1.783\% \pm 1.1\% &  9.814\% \pm 2.3\% &  9.258\% \pm 2.2\% &  9.345\% \pm 2.4\% &  1.035\% \pm 0.3\% & 86.08\% \pm 3.90\% \\
            0.1 & 2.682\% \pm 0.7\% & 12.999\% \pm 2.4\% & 12.221\% \pm 2.2\% & 10.645\% \pm 2.7\% &  3.972\% \pm 0.3\% & 86.09\% \pm 3.90\% \\
            0.2 & 3.288\% \pm 0.8\% & 15.018\% \pm 2.5\% & 14.423\% \pm 2.4\% & 11.620\% \pm 2.5\% &  7.639\% \pm 0.6\% & 86.13\% \pm 3.90\% \\
            0.3 & 3.813\% \pm 0.8\% & 16.521\% \pm 2.7\% & 16.198\% \pm 2.5\% & 11.646\% \pm 2.6\% & 11.242\% \pm 0.8\% & 86.21\% \pm 3.89\% \\
            0.4 & 4.366\% \pm 1.2\% & 19.635\% \pm 3.1\% & 19.442\% \pm 2.9\% & 11.876\% \pm 2.8\% & 14.744\% \pm 1.1\% & 86.32\% \pm 3.88\% \\
            0.5 & 5.091\% \pm 0.8\% & 20.848\% \pm 3.1\% & 19.974\% \pm 2.8\% & 13.578\% \pm 2.5\% & 18.237\% \pm 1.3\% & 86.46\% \pm 3.87\% \\
            0.6 & 5.457\% \pm 0.9\% & 24.459\% \pm 3.6\% & 24.875\% \pm 3.7\% & 13.331\% \pm 2.8\% & 21.777\% \pm 1.5\% & 86.63\% \pm 3.85\% \\
            0.7 & 6.496\% \pm 1.3\% & 25.153\% \pm 3.7\% & 26.143\% \pm 3.7\% & 14.096\% \pm 3.0\% & 25.466\% \pm 1.8\% & 86.83\% \pm 3.83\% \\
            0.8 & 7.001\% \pm 1.3\% & 27.023\% \pm 3.8\% & 26.910\% \pm 3.9\% & 14.457\% \pm 3.1\% & 29.411\% \pm 2.1\% & 87.06\% \pm 3.81\% \\
            0.9 & 7.689\% \pm 1.3\% & 27.328\% \pm 3.8\% & 28.393\% \pm 3.7\% & 14.656\% \pm 3.0\% & 33.673\% \pm 2.5\% & 87.32\% \pm 3.78\% \\
            1.0 & 7.624\% \pm 1.4\% & 29.294\% \pm 3.8\% & 28.469\% \pm 3.6\% & 14.923\% \pm 3.1\% & 38.294\% \pm 2.9\% & 87.61\% \pm 3.75\% \\
         \bottomrule
    \end{tabular}
	}
    \vspace{0.5em}
    \caption{Comparison of different methods trained in the presence of noise. Each model is trained at the same noise level used for testing.
    }
    \label{tab:noisy}
\end{table}

\begin{table}[h]
    \centering
    \caption*{ \textbf{Distribution shift}}
	\resizebox{1.0\textwidth}{!}{
    \begin{tabular}{lrrrrrr}
        \toprule
        \textbf{Contrast} 
        & \makecell{{\mmgunet} \\ (ours)}
        & \makecell{MFISNet \\ (smoothed)}
        & \makecell{MFISNet \\ (unsmoothed)}
        & WBENet
        & \makecell{Mod RecLin \\ (ours)}
        & Orig RecLin \\
        \midrule
        0.5 & $679.619\%\pm 121.9\%$ & $113.762\% \pm 17.2\%$ & $170.077\% \pm 16.0\%$ & $283.58\% \pm 44.5\%$ & $1.159\% \pm 0.1\%$ & $3.545\% \pm 0.3\%$ \\
        1.0 & $118.542\%\pm 81.2\%$ & $48.849\% \pm 9.8\%$ & $48.115\% \pm 15.3\%$ & $74.30\% \pm 16.8\%$ & $0.901\% \pm 0.1\%$ & $8.466\% \pm 2.0\%$ \\
        1.5 & $2.174\%\pm 3.9\%$ & $17.664\% \pm 7.4\%$ & $18.860\% \pm 5.8\%$ & $21.98\% \pm 9.8\%$ & $0.822\% \pm 0.1\%$ & $60.980\% \pm 8.4\%$ \\
        1.9 & $1.659\%\pm 0.8\%$ & $9.601\% \pm 2.8\%$ & $10.998\% \pm 2.7\%$ & $9.80\% \pm 3.5\%$ & $0.942\% \pm 0.2\%$ & $82.878\% \pm 5.0\%$ \\
        2.0 & $1.756\%\pm 0.8\%$ & $9.289\% \pm 2.2\%$ & $10.559\% \pm 2.2\%$ & $9.37\% \pm 2.6\%$ & $1.037\% \pm 0.3\%$ & $85.789\% \pm 4.1\%$ \\
        2.1 & $1.948\%\pm 0.7\%$ & $9.583\% \pm 1.8\%$ & $10.710\% \pm 1.8\%$ & $10.09\% \pm 2.0\%$ & $1.912\% \pm 12.2\%$ & $88.206\% \pm 3.5\%$ \\
        2.5 & $5.681\%\pm 4.7\%$ & $16.764\% \pm 3.9\%$ & $17.126\% \pm 3.6\%$ & $21.01\% \pm 5.2\%$ & $100.747\% \pm 127.4\%$ & $96.233\% \pm 3.2\%$ \\
        3.0 & $34.480\%\pm 27.7\%$ & $34.049\% \pm 5.6\%$ & $35.387\% \pm 7.1\%$ & $44.50\% \pm 8.3\%$ & $246.912\% \pm 75.9\%$ & $103.527\% \pm 2.8\%$ \\
        \bottomrule
    \end{tabular}
	}
    \vspace{0.5em}
    \caption{Comparison of different methods with the relative $\ell_2$ error, presented with the standard deviation over the relevant datasets. All models were trained on a dataset with maximum contrast $\norm{q}_\infty=2$.
    }
    \label{tab:ood}
\end{table}

\subsection{Forward model implementation}
\label{appendix:forward-model-code}
During training and inference, we evaluate $q \mapsto \mathcal{F}_{k_t}[q]$ and $v \mapsto \DF_{k_t}[q]^* v$ using a PDE solver based on the Hierarchical Poincar\'e-Steklov (HPS) method. This is a different solution strategy, with a different discretization, from the one used to create the dataset (\cref{appendix:dataset}); this choice was made to prevent any inverse crimes. %

HPS is a direct solver for linear elliptic boundary value problems. It uses a composite spectral collocation and a nested dissection strategy to achieve high-order error convergence and nearly-linear computational complexity \citep{martinsson_hierarchical_2015,gillman_spectrally_2015}. The highly-parallel structure of the algorithm makes it amenable to extremely fast evaluation on general-purpose graphics processing units \citep{melia_hardware_2025}. 
To enforce the  Sommerfeld radiation condition in \eqref{eq:pde_problem}, a boundary integral equation is used \citep{gillman_spectrally_2015}. This boundary integral equation is parameterized by single- and double-layer potential matrices generated using code from \citet{askham_chunkie_2024}
The HPS code is taken from \texttt{jaxhps} \citep{melia_jaxhps_2025}, a JAX implementation optimized for GPU use.

\subsection{Neural network implementation and hyperparameters}
\label{appendix:nn-hyperparams}
We present the choice of hyperparameters used in the different neural network methods. Every neural network is trained for 300 epochs, and, at the end of each training stage, the weights are chosen from the epoch with the lowest validation error.

\subsubsection{\hpscnn~(ours)}
The FYNet block uses the same architectural hyperparameters as in \citet{melia_multi-frequency_2025} in the $N_k=5$ setting: 1D kernel size 40; 2D kernel size 5; and 3 layers and 24 channels for each of the 1D and 2D blocks.
The 2D CNN components of the refinement block each contain 4 layers, 36 channels, and a kernel size of 5.
The optimization parameters selected were: a learning rate starting at $3\times 10^{-4}$ with minimum value $1\times 10^{-5}$, and weight decay of $1\times 10^{-3}$. Parameters were initialized using Kaiming He normalization \citep{he_delving_2015}.
Overall, our model has 965,822 parameters.

As for the forward model's settings, HPS solver is used with a quadtree of $L=3$ levels, with each leaf represented as a polynomial evaluated on a $p\times p$ Chebyshev grid where $p=16$.

\subsubsection{MFISNet-Refinement}
We use an implementation of MFISNet-Refinement and FYNet from \citet{melia_multi-frequency_2025}, along with the same architectural hyperparameters: 1D kernel size 40; 2D kernel size 5; and 3 layers and 24 channels for each of the 1D and 2D blocks. Slight modifications were made to the code to use unsmoothed intermediate training targets. The optimization parameters selected were: a learning rate starting at $3\times 10^{-4}$ with minimum value $1\times 10^{-5}$, and weight decay of $1\times 10^{-3}$. Parameters were initialized using Kaiming He normalization \citep{he_delving_2015}.
Overall, this model has 3,327,179 parameters.

\subsubsection{Wide-band Equivariant Network}
We adapt the implementation of Wide-band Equivariant Network from \citet{zhang_baseline_2025}. 
In particular, we focus on the uncompressed variant as this was shown to achieve lower errors than the compressed variant \citep{zhang_solving_2024}.
We modify their code to handle ten frequencies rather than three, as well as our differently laid-out dataset, and our fork of the repository is available at \url{https://github.com/oortsang/ISP_baseline_fork}.
We performed a hyperparameter grid search including the original hyperparameters outlined by \citet{zhang_solving_2024} and selected the following hyperparameters for the 2D CNN: 3 layers, 24 hidden channels, and a kernel size of 5. We trained with an initial learning rate of $10^{-5}$ and the same scheduler as in the original paper.

Overall, the uncompressed Wide-band Equivariant Network has 3,027,803 parameters as configured. (This is more than what \citet{zhang_solving_2024} report because we use ten frequencies instead of three, and our scattering potentials and measurements are discretized on a grid with dimensions $192\times 192$ rather than $80\times 80$.)

\subsection{Recursive linearization implementation}
\label{appendix:rl-impl}
We use a custom implementation of recursive linearization based on \citet{chen_recursive_1995,borges_high_2016} as well as a modified version.
We include several minor modifications in our ``original'' version of recursive linearization to slightly improve stability.
\citet{chen_recursive_1995} does not mention regularization, which has become standard for filtered back-projection \citep{fan_solving_2022}, whereas
\citet{borges_high_2016} implicitly regularizes updates by projecting scattering potentials down to a sine basis that grows with each frequency (to reflect the increased frequency content according to the diffraction limit).
Our implementation uses Tikhonov for every frequency to compensate for the expanded basis size, which we find to be necessary as reconstructions are very unstable without any regularization (the system being solved is poorly conditioned). This corresponds to the Levenberg-Marquardt algorithm \citep[see][chapter 18]{boyd_introduction_2018}. Our implementation also skips steps that would introduce NaNs into the solution or increase the measurement error. We refer to this version as ``original'' as we consider it sufficiently close to the two cited implementations, and the changes help this baseline; we also note \citet{borges_high_2016} already introduces several adjustments to the exact original statement of recursive linearization.

We also consider a custom modification to recursive linearization (which we call ``modified recursive linearization'' or ``Modified RecLin''), outlined in \cref{alg:mod-rl}, intended to better handle the relatively small number of steps. Whereas \citet{borges_high_2016} uses hundreds of frequencies, our setting only includes ten frequencies; it seems reasonable to compensate by using additional computation. Instead of taking a single Levenberg-Marquardt (or Gauss-Newton) step per frequency, we perform multiple Levenberg-Marquardt steps (using additional steps is the most important for the initial low-frequency estimate). We find that this greatly improves reliability and reconstruction quality. 

For both versions of recursive linearization, we select hyperparameters using our dataset's validation set, where all potentials have maximum contrast $\norm{q}_\infty=2$. All additional testing is performed using the same hyperparameters, which is likely a contributing factor in why Modified RecLin performs so poorly at higher contrasts. See \cref{tab:rl-params} for the selected hyperparameters.

\SetKw{Continue}{continue}
\SetKw{Break}{break}
\begin{algorithm}[t]
\SetAlgoLined
\DontPrintSemicolon
\KwData{Measurement data $d_{k_1}, d_{k_2}, \dots, d_{k_{N_k}}$, number of iterations per frequency $n_{k_1},n_{k_2},\dots,n_{k_{N_k}}$, and regularization parameters $\varepsilon_{k_1},\varepsilon_{k_2},\dots,\varepsilon_{k_{N_k}}$}
\caption{Modified recursive linearization}
\label{alg:mod-rl}
\vspace{8pt}
$\hat q_{k_0} \leftarrow 0$ \\
\For{$t=1,2,3,\dots,N_{k}$}{
    $\hat q_\mathrm{prev} \leftarrow \hat q_{k_{t-1}}$ \\
    \For{$\ell=1,\dots,{n_{k_t}}$}{
        $\gme \leftarrow \DF_{k_t} [\hat q_\mathrm{prev}]^*
    (d_{k_t} - \mathcal F_{k_t} [\hat q_\mathrm{prev}] )$  \tcc*[f]{back-projection step}\\
        $\delta \hat q \leftarrow 
        (\DF_{k_t} [\hat q_\mathrm{prev}]^*\DF_{k_t} [\hat q_\mathrm{prev}] 
        + \varepsilon_{k_t} I)^{-1}
        \gme
        $ \tcc*[f]{filtering step}\\
        \If{
            $\norm{d_{k_t} - \mathcal F_{k_t}[\hat q_\mathrm{prev} + \delta \hat q ] }_2
            \le
            \norm{d_{k_t} - \mathcal F_{k_t}[\hat q_\mathrm{prev}]}_2
            $
        } {
            $\hat q_\mathrm{next} \leftarrow \hat q_\mathrm{prev} + \delta \hat q$
            \tcc*[f]{step if it improves measurement error}
        } \Else {
            $\hat q_\mathrm{next} \leftarrow \hat q_\mathrm{prev}$ \\
            \Break %
        }
        $\hat q_\mathrm{prev} \leftarrow \hat q_\mathrm{next}$ \tcc*[f]{update inner loop variable} \\
    }
    $\hat q_{k_t} \leftarrow \hat q_\mathrm{next}$
}
\Return{$\hat q_{k_{N_k}}$}
\end{algorithm}

\begin{table}[h]
    \centering
    \begin{tabular}{LRR}
        \toprule
            \textbf{Parameter} & \text{Mod RecLin} & \text{Orig RecLin} \\
        \midrule
            n_{k_1} & 10 & 1 \\
            n_{k_2},\dots,n_{k_{N_k}} & 2 & 1 \\
            \varepsilon_{k_1} & 10^{-4} & 10^{-3} \\
            \varepsilon_{k_2},\dots,\varepsilon_{k_{N_k}} & 10^{-2} & 10^{-1} \\
         \bottomrule \\
    \end{tabular}
    \caption{
        Parameters selected for each variant of recursive linearization.
    }
    \label{tab:rl-params}
\end{table}

\subsection{Forward model equations}
\label{appendix:forward-model-equations}
We give integral equations for the forward model and its adjoint derivative operator to help illustrate their behavior, particularly the nonlinear dependence on $q$. We will also show why the forward model is approximately linear in the low-frequency or low-contrast regimes.

First, we define several operators based on the Green's function, $G_k(x-x')= -\frac{i}{4} H_0^{(1)}(k\lVert x-x'\rVert_2)$, which is the zeroth order Hankel function of the first kind \citep{borges_high_2016} (we use a flipped sign convention as compared with \cite{borges_high_2016}). With slight abuse of notation,
\begin{align}
(G_k v)(x)
&= \int_\Omega G_k(x-x') v(x') dx' \\
(M_k v)(r)
&= \int_\Omega G_k(Rr-x') v(x') dx'.
\end{align}
$G_k$ maps a field $v$ from $\Omega$ to $\Omega$, while $M_k$ is the far-field measurement operator mapping $v$ from scattering domain $\Omega$ to the receivers at radius $R$. Additionally, we indicate point-wise multiplication as
\begin{equation}
(\diag (q) v)(x) = q(x)v(x).
\end{equation}

In this notation, the Lippmann-Schwinger equations for source direction $s$, in the near- and far-field cases, respectively, are
\begin{align}
\usc^{s} &= k^2 G_k \diag (q) (\uin^{s} + \usc^{s}) \label{eq:ls-near} \\
\usc^{s, \text{far}} &= k^2 M_k \diag (q) (\uin^{s} + \usc^{s}) \label{eq:ls-far},
\intertext{which can be rearranged to give the forward model for source $s$ as}
(\mathcal F_{k, s}[q])
&= \usc^{s, \text{far}} \\
&= k^2 M_k (I-k^2 \diag (q) G_k)^{-1} \diag (q) \uin^{s}.
\label{eq:ls-fwd-model}
\end{align}
Note that when $k$ or $q$ are small in magnitude, the factor $(I-k^2 \diag (q) G_k)^{-1} \approx I$, which means the forward model behaves nearly linearly in those regimes.

This particular linear approximation does not hold in general, so recursive linearization \citep{chen_recursive_1995} proposes to approximate the forward model using a first-order Taylor expansion:
\begin{align}
\mathcal F_k[q+\delta q]
&\approx \mathcal F_k[q] + \DF_k[q] \delta q.
\end{align}
where $\DF_k[q]$ is the Fr\'echet derivative of $\mathcal F_k$ at $q$.
For a given source direction $s$, the derivative and its adjoint, acting on $\delta q$ and $\xi_s$, respectively, are given by
\begin{align}
\DF_{k, s} [q] \delta q &= k^2 M_k (I-k^2 \diag (q) G_k)^{-1} \diag (\delta q) \utot^s 
\label{eq:derivative-operator} \\
\DF_{k, s} [q]^* \xi_s &= \text{Re} \left[ k^2 \diag (\overline{\utot^s}) (I-k^2 G_k^* \diag (q))^{-1} M_k^* \xi_s \right]
\label{eq:adjoint-derivative-operator}
\end{align}
where $\utot^s=\uin^s+\usc^s$ is the total wave field and $\overline{(\cdot)}$ indicates element-wise complex conjugation. The adjoint derivative operator uses only the real component since the scattering potential $q$ is real-valued in our setting. We omit the derivation, but these equations can be derived from the PDEs given by \citet[Theorems 3.1, 3.2]{borges_high_2016} or by applying the adjoint state method (see \cite{bradley_pde_2024} for a nice tutorial).

Even though we are interested in $\DF_{k} [q]$ and $\DF_{k} [q]^*$ for the sake of linearizing the forward model, it is important to remark that these operators themselves depend on $q$ in a nonlinear manner (i.e., the maps $v\mapsto\DF_k[q]v$ and $v\mapsto\DF_k[q]^* v$ are linear \emph{in $v$ but not in $q$}). This is because $q$ appears inside the matrix inverses. So, even if $\delta q$ or $\xi_s$ are small, the nonlinear dependence on $q$ is not negligible. (If $k$ or $q$ were small, we could stick with the previous linearization.)

The functional form of the operators' dependence on $q$ also suggests that it may be challenging to model the derivative operators using a neural network counterpart (in a way that holds for large $\norm{q}_\infty$). This is part of why we think our method's use of the forward model's derivative is important to the method's success.
In earlier investigations, we attempted to use a neural network to replace the mapping $(q, d_k, \xi) \mapsto \DF_{k}[q]^* \xi$, where $\xi=d_k-\mathcal F_k[q]$ is computed using the true forward model. The architecture was based off MFISNet-Refinement's architecture, except that each refinement block's FYNet component received $(d_k, \mathcal F_k[q], d_k-\mathcal F_k[q])$ as inputs instead of just $d_k$.
However, this alternate method saw only modest improvements over MFISNet-Refinement \citep{melia_multi-frequency_2025}.
We found much larger improvements after switching to applying the true operator $\DF_{k}[q]^*$, which suggests that in general this operator is likely more difficult to express than $F_k^*$.

\end{document}